\documentclass[prd,twocolumn,showpacs,superscriptaddress]{revtex4}
\usepackage{graphicx}
\usepackage{amsmath}

\DeclareMathOperator{\Tr}{Tr}

\begin{document}

\title{Threshold Resummation Effects in Direct Top Quark Production at Hadron Colliders}
\author{Li Lin Yang}
\email{llyang@pku.edu.cn}
\author{Chong Sheng Li}
\email{csli@pku.edu.cn}
\author{Yang Gao}
\affiliation{Department of Physics, Peking University, Beijing 100871, China}
\author{Jian Jun Liu}
\affiliation{Center for High Energy Physics and Department of Physics, Tsinghua University, Beijing 100084, China}
\date{\today}

\begin{abstract}
  We investigate the threshold-enhanced QCD corrections to the cross sections for direct
  top quark productions induced by model-independent flavor changing neutral current
  couplings at hadron colliders. We use the soft-collinear effective theory to describe
  the incoming massless partons and use the heavy quark effective theory to treat the top
  quark. Then we construct the flavor changing operator based on the above effective
  theories, and resum the large logarithms near threshold arising from soft gluon
  emission. Our results show that the resummed QCD corrections further enhance the
  next-to-leading order cross sections significantly. Moreover, the resummation effects
  vastly reduce the dependence of the cross sections on the renormalization and
  factorization scales, especially in cases where the next-to-leading order results
  behave worse than the leading order results. Our results are more sensitive to the new
  physics effects. If signals of direct top quark production are found in future
  experiments, it is more appropriate to use our results as the theoretical inputs for
  extracting the anomalous couplings.
\end{abstract}


\maketitle

%%ALIAS=young=Phys.Rev.D56.5725,Phys.Rev.D58.073008%%
%%ALIAS=ljj=Phys.Rev.D72.074018%%
%%ALIAS=scet=Phys.Rev.D63.014006,Phys.Rev.D63.114020,Phys.Lett.B516.134%%
The production and decay of top quark through FCNC couplings are very sensitive to new
physics effects, which have been extensively studied in some new physics models in the
literature (see Ref.~\cite{Acta.Phys.Polon.B35.2695} and references there in). Since we
do not know which type of new physics will be responsible for a future deviation from the
SM predictions, it is necessary to study the top quark FCNC processes in a model
independent way by an effective Lagrangian. For $t$-$q$-$g$ anomalous couplings, the
direct top quark production is the most sensitive process. The analysis based on the
leading order (LO) cross sections~\cite{young} suggests that the anomalous couplings can
be detected down to 0.019~TeV$^{-1}$ for $q=u$ and 0.062~TeV$^{-1}$ for $q=c$ at the
Tevatron Run 2. Studies with a fast detector simulation for ATLAS indicate a similar
reach at the LHC~\cite{J.Phys.G31.N1}.

As we know, the LO cross sections for processes at hadron colliders suffer from large
uncertainties due to the arbitrary choice of the renormalization scale ($\mu_r$) and
factorization scale ($\mu_f$), and are not sufficient for the extraction of the anomalous
couplings from experiments. In general, a next-to-leading order (NLO) QCD calculation is
capable to reduce the scale dependence significantly. But it is not the case for direct
top quark production considered here. In Ref.~\cite{ljj}, the cross sections for the
direct top production are calculated at NLO in QCD. Their results showed that the NLO QCD
corrections reduced the scale dependence of the total cross sections at the Tevatron Run
2. However, the scale dependence was not improved at the LHC, and even became worse in
the region $\mu_r = \mu_f < m_t$. It was pointed out~\cite{ljj} that the scale dependence
comes mainly from the terms proportional to $\delta(1-z)$ in the partonic cross sections,
and higher order effects are necessary to further reduce such scale dependence in the
case of $\mu_r=\mu_f$ at the LHC. In this paper, we report that the threshold resummation
effects can remarkably improve the scale dependence of the cross sections.

We consider the process $A + B \to g(k_1) + q(k_2) \to t + X$, where $A$, $B$ are the
colliding hadrons. At the parton level, the process is induced by the effective
Lagrangian
\begin{equation}
  \label{eq:lfull}
  \mathcal{L}_{\text{eff}} = - g_s \sum_{q=u,c} \frac{\kappa_{tq}^g}{\Lambda} \bar{t}
  \sigma^{\mu\nu} T^a (f_{tq}^g + ih_{tq}^g\gamma_5) q G_{\mu\nu}^a + \text{h.c.},
\end{equation}
where $\Lambda$ is the new physics scale, $T^a$ are the Gell-Mann matrices satisfying
$\Tr(T^aT^b) = \delta^{ab}/2$, $G_{\mu\nu}^a$ are the field strength tensors of gluon
fields. $\kappa$ is normalized to be real and positive and $f$, $h$ to be complex numbers
satisfying $|f|^2+|h|^2=1$.

The NLO results contain terms like $\left(\frac{\ln(1-z)}{1-z}\right)_+$ and
$\frac{1}{(1-z)_+}$, which are singular near the kinematical threshold $z \to 1$. Here $z
= Q^2/s$, $Q^2=m_t^2$ and $s=(k_1+k_2)^2$. Physically, these singular terms originate
from the emission of soft or collinear gluons. The soft-collinear effective theory
(SCET)~\cite{scet} is a natural framework to deal with the physics of soft and collinear
gluons and quarks. In the following, we will use our NLO results to derive a threshold
resummation formula for direct top quark production using the SCET.

We introduce the parameter $\lambda^2 \sim 1-z \gg \Lambda_{\text{QCD}}/Q$ in
SCET$_{\text{I}}$ and match the full theory operator (\ref{eq:lfull}) onto the operator
in the effective theory at leading order in $\lambda$. This step is similar as the ones
for the deep inelastic scattering (DIS) process in Ref.~\cite{Phys.Rev.D68.114019} and
the Drell-Yan process in Ref.~\cite{Phys.Rev.D72.054016}. However, there is a non-trivial
point in our problem concerning the top quark field. In general, the top quark field can
not be described in the SCET. However, in the threshold region $z \lesssim 1$, the top
quark is nearly at rest, which implies that its momentum can be written as $q=m_tv+k$,
where the residue momentum $k \sim Q\lambda^2$ comes from its interactions with soft
gluons. This situation is much similar as the bottom quark in $B$ decays. Therefore, we
can treat the top quark field using the heavy quark effective theory (HQET) by a
systematic expansion in $1/m_t$. Thus the effective operator can be written as
\begin{align}
  \label{eq:leff}
  \mathcal{L}_{gq} &= g_s \frac{\kappa}{\Lambda} \bar{t}_v \Gamma^{\mu}
  \mathcal{B}_{n\mu} C_{gq}(Q^2,\mu) W_{\bar{n}}^{\dagger} \xi_{\bar{n}} \nonumber
  \\
  &\equiv g_s \frac{\kappa}{\Lambda} C_{gq}(Q^2,\mu) \mathcal{T}_{gq},
\end{align}
where $\Gamma^{\mu}=\frac{1}{2}(f+ih\gamma_5)n_{\nu}\sigma^{\mu\nu}$, $t_v$ is the heavy
quark field in HQET describing the top quark, $\xi_{\bar{n}}$ is the collinear quark
field, $\mathcal{B}_n^\mu$ is related to the field strength tensor of the collinear gluon
field in SCET and $W_{\bar{n}}$ denotes a Wilson line which are required to ensure gauge
invariance of the operator. $n$ and $\bar{n}$ are two light-cone vectors satisfying
$n^2=\bar{n}^2=0$ and $n\cdot\bar{n}=2$. $C_{gq}$ is the matching coefficient which are
obviously equal to unity at tree level. We can determine the $\mathcal{O}(\alpha_s)$
matching conditions by evaluating the on-shell matrix elements of operators in both full
theory and SCET. In dimensional regularization (DR), the facts that the IR structure of
the full theory and the effective theory is identical and the on-shell integrals are
scaleless and vanish in SCET and HQET imply that the IR divergence of the full theory is
just the negative of the UV divergence of SCET. Thus, from the $\mathcal{O}(\alpha_s)$
virtual corrections, we can obtain the NLO matching coefficient and the anomalous
dimension of the effective operator $\mathcal{T}_{gq}$:
\begin{align}
  C_{gq}(Q^2,\mu) &= 1 + \frac{\alpha_{s}}{12\pi} \left[ -12\ln\frac{\mu^2}{Q^2}
    -\frac{13}{2}\ln^2\frac{\mu^2}{Q^2} \right. \nonumber
  \\
  &\qquad \qquad \left. - 23 + \frac{55\pi^2}{12} \right],
  \\
  \label{eq:gamma1}
  \gamma_1(\mu) &= \frac{\alpha_{s}}{6\pi} \left[ 13\ln\frac{\mu^2}{Q^2} + (6\beta_0+10)
  \right],
\end{align}
where $\beta_0 = 11/4 - n_f/6$.

Next, we perform the usual field redefinition to decouple the collinear and usoft fields
and match the operator to SCET$_{\text{II}}$. This matching is trivial: the matching
coefficient is unity and the anomalous dimension is the same as Eq.~(\ref{eq:gamma1}) and
we will still denote the new operator as $\mathcal{T}_{gq}$. Now we calculate the cross
section in SCET$_\text{II}$ with $\mathcal{T}_{gq}(\mu)$ at the scale $\mu \sim
Q\lambda^2$, and then match the result onto the product of two PDFs, which means
\begin{align}
  \label{eq:mf}
  \sigma_{\text{SCET}} &= \mathcal{M}(z,\mu) \otimes \left[ f_g(z,\mu) \otimes f_q(z,\mu)
  \right] \nonumber
  \\
  &\equiv \mathcal{M}(z,\mu) \otimes \mathcal{F}(z,\mu),
\end{align}
where $\mathcal{M}(z,\mu)$ is the matching coefficient. Since the virtual corrections in
SCET vanish at NLO in DR, $\sigma_{\text{SCET}}$ can be obtained by calculating the real
gluon emission diagrams in SCET and taking into account the renormalization of the
operator. The result is
\begin{multline}
  \label{eq:mm}
  \frac{\hat{\sigma}_{\text{SCET}}}{\hat{\sigma}_0} = \frac{\alpha_{s}}{6\pi} \left\{
    \frac{1}{\epsilon} \left[ -\frac{26}{(1-z)_+} - (6\beta_0+6) \delta(1-z) \right]
  \right.
  \\
  + \left[ 8 - \frac{13\pi^2}{4} + 4\ln\frac{\mu^2}{Q^2} + \frac{13}{2}
    \ln^2\frac{\mu^2}{Q^2} \right] \delta(1-z)
  \\
  - \frac{26}{(1-z)_+} \ln\frac{\mu^2}{Q^2} - \frac{8}{(1-z)_+} + 52 \left(
    \frac{\ln(1-z)}{1-z} \right)_+
  \\
  \left. - \frac{27}{2} \ln(1-z) - \frac{28\ln{z}}{1-z} + 7 \right\},
\end{multline}
where $\hat{\sigma}_0=(8\pi^2/3)\alpha_s(\kappa/\Lambda)^2z$. Matching the cross section
onto the product of two PDFs, all the poles are cancelled by the renormalization of the
PDFs,
\begin{equation*}
  Z_{\mathcal{F}}(z) = \delta(1-z) + \frac{\alpha_{s}}{6\pi} \frac{1}{\epsilon} \left[
    \frac{26}{(1-z)_+} + (6\beta_0+6) \delta(1-z) \right].
\end{equation*}
The resulting matching coeffcient is given by
\begin{multline}
  \mathcal{M}(z,\mu) = \frac{\alpha_{s}}{6\pi} \left\{ \left[ 8 - \frac{13\pi^2}{4} +
      4\ln\frac{\mu^2}{Q^2} + \frac{13}{2}\ln^2\frac{\mu^2}{Q^2} \right] \right.
  \\
  \times \delta(1-z) - \frac{26}{(1-z)_+} \ln\frac{\mu^2}{Q^2} - \frac{8}{(1-z)_+}
  \\
  + 52 \left( \frac{\ln(1-z)}{1-z} \right)_+ \left. - \frac{27}{2}\ln(1-z) -
    \frac{28\ln{z}}{1-z} + 7 \right\}.
\end{multline}
To factorize the convolution in Eq.~(\ref{eq:mf}), we make Mellin transformation into
moment space and neglect terms that vanish in the limit $N \to \infty$. The moment of
$\mathcal{M}$ is
\begin{equation}
  \mathcal{M}_N(\mu) = \frac{\alpha_s}{6\pi} \left[ 8 + \frac{13\pi^2}{12} +
    26\ln^2\frac{\bar N \mu}{Q} + 8\ln\frac{\bar{N}\mu}{Q} \right],
\end{equation}
where $\bar{N}=Ne^{\gamma_E}$ and $\gamma_E$ is the Euler constant. To avoid large logs
we choose the matching scale $\mu \sim Q/\bar N$. The running of the moments of
$\mathcal{F}$ is governed by it anomalous dimension
\begin{equation}
  \label{eq:gamma2}
  \gamma_2(\mu) = \frac{\alpha_s}{3\pi} \left[ -26\ln\bar{N} + 6\beta_0 + 6 \right].
\end{equation}

Combining the above results, we can write down the resummed cross section in moment space
\begin{align}
  \label{eq:scet}
  \sigma_N^{\text{SCET}} &= \sigma_0(\mu) \left| C_{gq}(Q^2,\mu) \right|^2 \left[ 1 +
    \mathcal{M}_N(\mu) \right] \mathcal{F}_N(\mu) \nonumber
  \\
  &= \sigma_0(\mu_r) \left| C_{gq}(Q^2,\mu_r) \right|^2 e^{I_1} \nonumber
  \\
  &\quad \times \left[ 1 + \mathcal{M}_N(Q/\bar{N}) \right] e^{I_2} \mathcal{F}_N(\mu_f),
\end{align}
with
\begin{equation}
  \label{eq:int}
  I_1 = \int_{Q/\bar{N}}^{\mu_r} \frac{d\mu}{\mu} 2\gamma_1(\mu), \qquad
  I_2 = \int_{\mu_f}^{Q/\bar{N}} \frac{d\mu}{\mu} \gamma_2(\mu),
\end{equation}
where $\mu_r$ and $\mu_f$ correspond to the renomalization and factorization scales in
the full theory, respectively, which are expected to satisfy $\mu_r \sim \mu_f \sim Q \gg
Q/\bar{N} \gg \Lambda_{\text{QCD}}$.

To reach the accuracy of next-to-leading logarithms (NLL) in the exponents, we need the
coefficient of $\ln(\mu^2/Q^2)$ in $\gamma_1(\mu)$ and that of $\ln\bar{N}$ in
$\gamma_2(\mu)$ up to two-loop. It can be shown that $\gamma_1$ and $\gamma_2$ have the
form~\cite{Phys.Rev.D68.114019}
\begin{align}
  \gamma_1(\mu) &= A_1(\alpha_s) \ln\frac{\mu^2}{Q^2} + A_0(\alpha_s), \nonumber
  \\
  \gamma_2(\mu) &= B_1(\alpha_s) \ln\bar{N} + B_0(\alpha_s),
\end{align}
and $4 A_1(\alpha_s) = -B_1(\alpha_s)$.
%%ALIAS=dglap=Nucl.Phys.B175.27,Phys.Lett.B97.437%%
Thus we can extract $A_1$ from the information of $B_1$ at the two-loop level. $B_1$ is
related to the coefficients of $1/(1-z)_+$ in the DGLAP splitting kernels. Using the
two-loop splitting functions~\cite{dglap} together with Eq.~(\ref{eq:gamma1}) and
(\ref{eq:gamma2}), we can get the coefficients
\begin{align}
  A_1^{(1)} &= -\frac{1}{4} B_1^{(1)} = \frac{13}{6}, \nonumber
  \\
  A_1^{(2)} &= -\frac{1}{4} B_1^{(2)} = \frac{A_1^{(1)}}{2} \left[ C_A \left(
      \frac{67}{18} - \frac{\pi^2}{6} \right) - \frac{5}{9}n_f \right], \nonumber
  \\
  A_0^{(1)} &= \beta_0 + \frac{5}{3}, \quad B_0^{(1)} = 2\beta_0 + 2,
\end{align}
where $A_1=\sum(\alpha_{s}/\pi)^{n}A_1^{(n)}$ and similar for $B_1$, $A_0$ and $B_0$.
Note that the coefficients $A_1$ and $B_1$ only depend on the initial states and can be
related to the quark-quark coefficients and the gluon-gluon coefficients in a
straightforward way. For example, $A_1^{(1)} = (A_q^{(1)}+A_g^{(1)})/2 \equiv
(C_F+C_A)/2$.

Now we can evaluate the integrals in Eq.~(\ref{eq:int}) using the two-loop evolution of
$\alpha_s$ in the $\overline{\text{MS}}$ scheme. Keeping only terms up to NLL in the
exponents in Eq.~(\ref{eq:scet}), we obtain
\begin{align}
  I_1 + I_2 &= \ln{N} g^{(1)}(\beta_0\alpha_s(\mu_r)\ln{N}/\pi) \nonumber
  \\
  + &g^{(2)}(\beta_0\alpha_s(\mu_r)\ln{N}/\pi) + \mathcal{O}(\alpha_s(\alpha_s\ln{N})^k),
  \\
  g^{(1)}(\lambda) &= \frac{A_1^{(1)}}{\beta_0\lambda} [ 2\lambda + (1-2\lambda)
  \ln(1-2\lambda) ],
  \\
  g^{(2)}(\lambda) &= - \frac{2A_1^{(1)}\gamma_E}{\beta_0} \ln(1-2\lambda) +
  \frac{A_1^{(1)}\beta_1}{\beta_0^3} \left[ 2\lambda + \ln(1-2\lambda) \right. \nonumber
  \\
  &\left. + \frac{1}{2}\ln^2(1-2\lambda) \right] -\frac{A_1^{(2)}}{\beta_0^2} [ 2\lambda
  + \ln(1-2\lambda) ] \nonumber
  \\
  &- \frac{A_1^{(1)}}{\beta_0} \ln(1-2\lambda) \ln\frac{\mu_r^2}{Q^2} -
  \frac{A_1^{(1)}}{\beta_0} 2\lambda \ln\frac{\mu_r^2}{\mu_f^2} \nonumber
  \\
  &\qquad \qquad + \frac{B_0^{(1)}-2A_0^{(1)}}{2\beta_0} \ln(1-2\lambda),
\end{align}
which are the same as the results obtained within full QCD \cite{disdy,toppair} except
the coefficients $A$ and $B$. The NLL cross section in moment space is then given by
\begin{equation*}
  \sigma^{\text{NLL}}_N = \sigma_0 C(\alpha_s,Q^2,\mu_r,\mu_f) \exp
  \left(g^{(1)}\ln{N}+g^{(2)}\right) \mathcal{F}_N(\mu_f),
\end{equation*}
where $C$ represents the contributions from $C_{gq}$, $\mathcal{M}_N$, as well as the
terms neglected in the exponents. To obtain the physical cross section, we perform the
inverse Mellin transformation back to the $x$-space
\begin{equation}
  \sigma^{\text{NLL}}(\tau) = \frac{1}{2\pi i} \int_C dN \tau^{-N} \sigma^{\text{NLL}}_N.
\end{equation}
We use the tricks introduced in Ref.~\cite{Phys.Rev.D66.014011} to evaluate the
$N$-integral numerically. Finally, The resummed cross section at NLL accuracy is defined
to be the NLL cross section plus the remaining terms in the NLO result which are not
resummed, i.e.,
\begin{equation*}
  \sigma^{\text{Resum}} = \sigma^{\text{NLL}} + \sigma^{\text{NLO}} -
  \sigma^{\text{NLL}} \bigg|_{\alpha_s=0} - \alpha_s \left(
    \frac{\partial\sigma^{\text{NLL}}}{\partial\alpha_s} \right)_{\alpha_s=0}.
\end{equation*}

We now present the numerical results of the NLL threshold resummation for the direct top
quark production at the Tevatron Run 2 and the LHC, respectively. In our numerical
calculations, we take the top quark mass $m_t = 174.3$~GeV~\cite{Phys.Lett.B592.1}, which
is different from the value 178.0~GeV used in Ref.~\cite{ljj}.
% The anomalous couplings, which appear in the expressions of the cross sections as
% quadratic factors, are chosen to be $\kappa/\Lambda=0.01$~TeV$^{-1}$.
We use both the CTEQ6 PDFs~\cite{hep-ph/0201195} and the MRST2004
PDFs~\cite{hep-ph/0410230} to evaluate the total cross sections.

In Table~\ref{tab:cs}, we list the LO, NLO and resummed cross sections for the
renormalization and factorization scales $\mu_r=\mu_f=m_t$ using the two PDF sets. It can
be seen that the threshold resummation effects further enhance the NLO cross sections by
about 30\%. When the resummed results are compared with the LO ones, the enhancement can
even reach 100\% in most of cases. We also note that the discrepancies between the
different PDF sets are somewhat larger in the case of resummed than the case of LO or
NLO, especially at the LHC. This may be due to the different fitting methods used in the
PDF sets. We believe that when the LHC data be used in the global fitting of the PDFs,
these discrepancies will be improved.

\begin{table*}[ht!]
\setlength{\tabcolsep}{1em}
\begin{tabular}{c|c|ccc|ccc}
  \hline \hline
  & & \multicolumn{3}{c|}{LHC $\left(\frac{\kappa/\Lambda}{0.01\text{TeV}^{-1}}\right)^2$pb} &
  \multicolumn{3}{c}{Tevatron $\left(\frac{\kappa/\Lambda}{0.01\text{TeV}^{-1}}\right)^2$fb} 
  \\ \cline{3-8}
  \raisebox{2ex}[0cm][0cm]{subprocess} & \raisebox{2ex}[0cm][0cm]{PDF} & LO & NLO & Resum
  & LO & NLO & Resum
  \\ \hline
  & CTEQ & 12.9 & 17.0 & 23.7 & 268 & 425 & 547
  \\
  \raisebox{3ex}[0cm][0cm]{$gu \to t$} & MRST & 12.2 & 16.3 & 19.5 & 262 & 426 & 520
  \\ \hline
  & CTEQ & 1.71 & 2.53 & 3.71 & 13.1 & 28.1 & 38.2
  \\
  \raisebox{3ex}[0cm][0cm]{$gc \to t$} & MRST & 1.68 & 2.38 & 2.92 & 17.0 & 30.3 & 38.6
  \\
  \hline \hline
\end{tabular}
\caption{\label{tab:cs}The LO, NLO and resummed cross sections for direct top quark
  production at the LHC and Tevatron Run 2. Here $\mu_r=\mu_f=m_t$.}
\end{table*}

\begin{figure}[ht!]
  \includegraphics[width=0.23\textwidth]{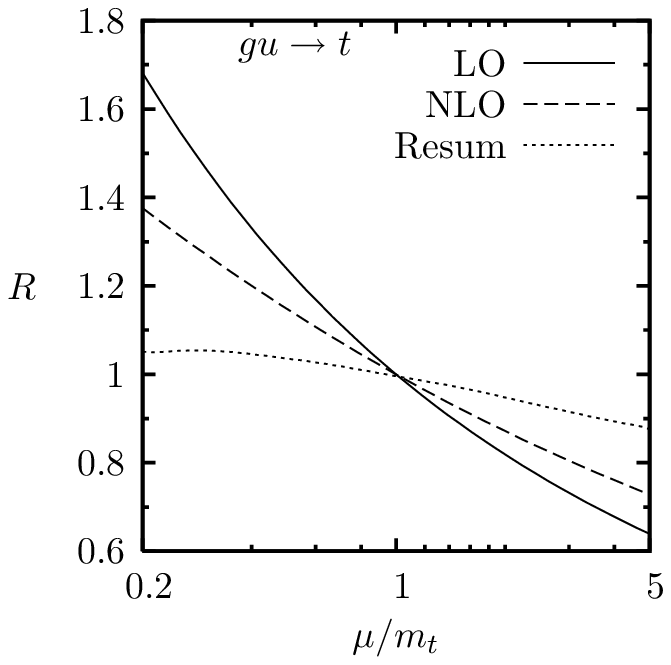}
  \includegraphics[width=0.23\textwidth]{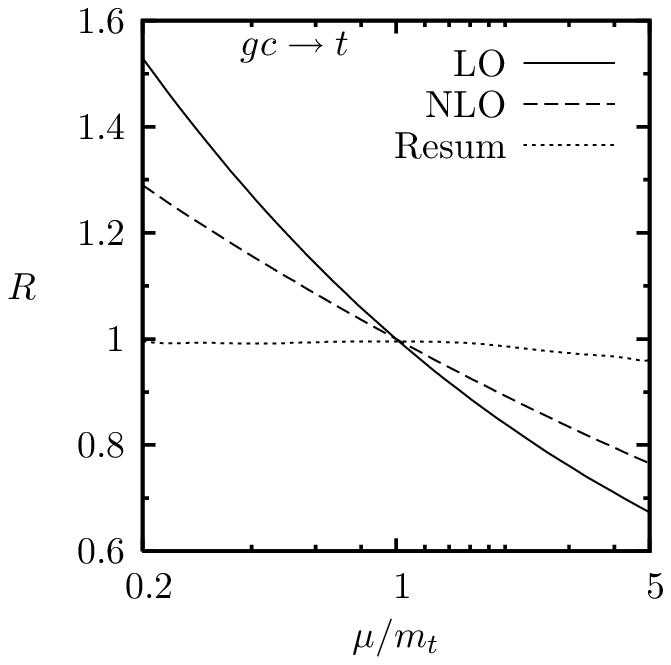}
  \caption{\label{fig:tev_cteq}The ratio $R$ as functions of the renormalization and
    factorization scales at the Tevatron Run 2 for subprocesses $gu \to t$ (left) and
    $gc \to t$ (right) using CTEQ6 PDFs.}
\end{figure}

\begin{figure}[ht!]
  \includegraphics[width=0.23\textwidth]{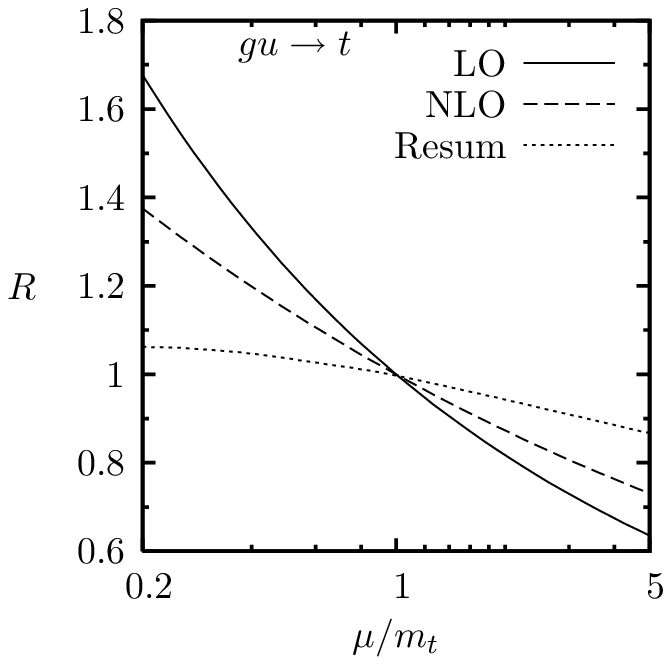}
  \includegraphics[width=0.23\textwidth]{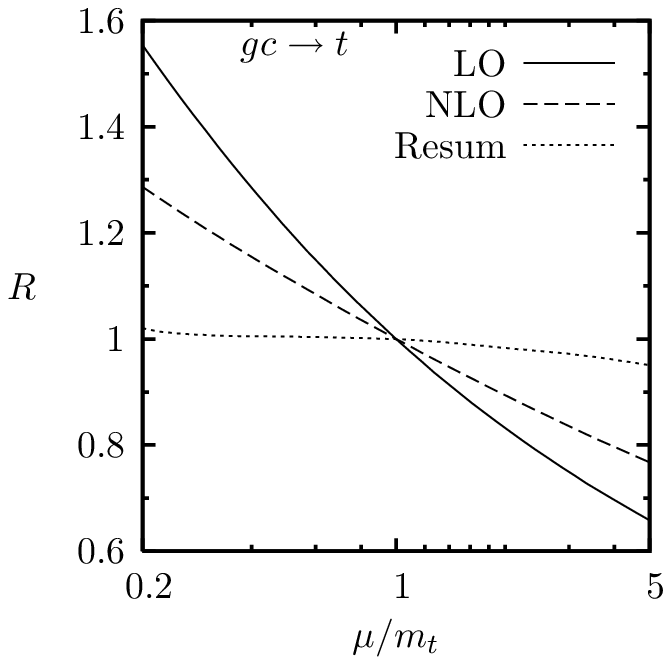}
  \caption{\label{fig:tev_mrst}The ratio $R$ as functions of the renormalization and
    factorization scales at the Tevatron Run 2 for subprocesses $gu \to t$ (left) and
    $gc \to t$ (right) using MRST2004 PDFs.}
\end{figure}

Despite the uncertainties from the PDFs, we can estimate the theoretical uncertainties by
investigating the sensitivity of the results to the renomalization and factorization
scales. We define $R$ as the ratio of the cross sections (LO, NLO, resummed) to their
values at central scale, $\mu_r=\mu_f=m_t$, always assuming $\mu_r=\mu_f=\mu$ for
simplicity. Figure~\ref{fig:tev_cteq} and Figure~\ref{fig:tev_mrst} show the ratio $R$ as
functions of the renormalization and factorization scales for subprocesses $gu \to t$ and
$gc \to t$ at the Tevatron Run 2 using the two PDF sets, respectively. We can find that,
for both subprocesses, the threshold resummation effects further decrease the scale
dependence of the NLO cross sections remarkably. For example, the variations of the cross
sections in the region $m_t/2 \leq \mu \leq 2m_t$, for subprocess $gu \to t$, are about
28\% for NLO cross section and only 9\% for resummed one. In the case of $gc \to t$, they
are about 11\% and 2\%, respectively.

\begin{figure}[ht!]
  \includegraphics[width=0.23\textwidth]{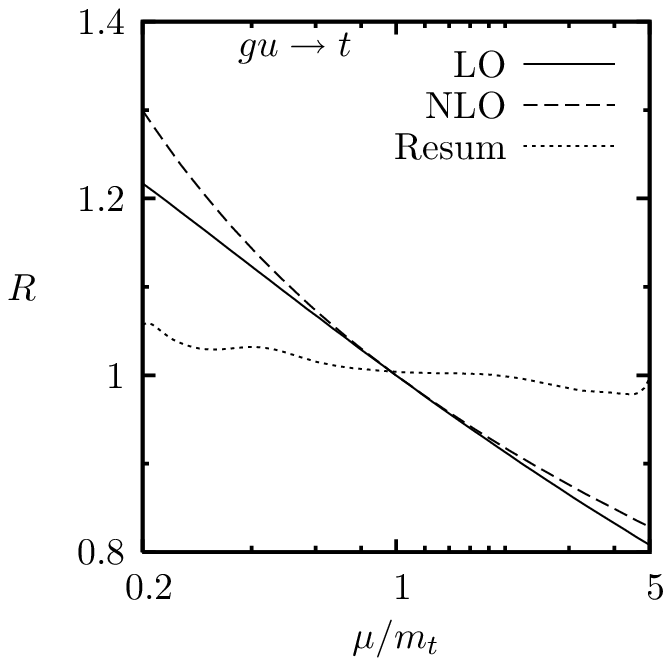}
  \includegraphics[width=0.23\textwidth]{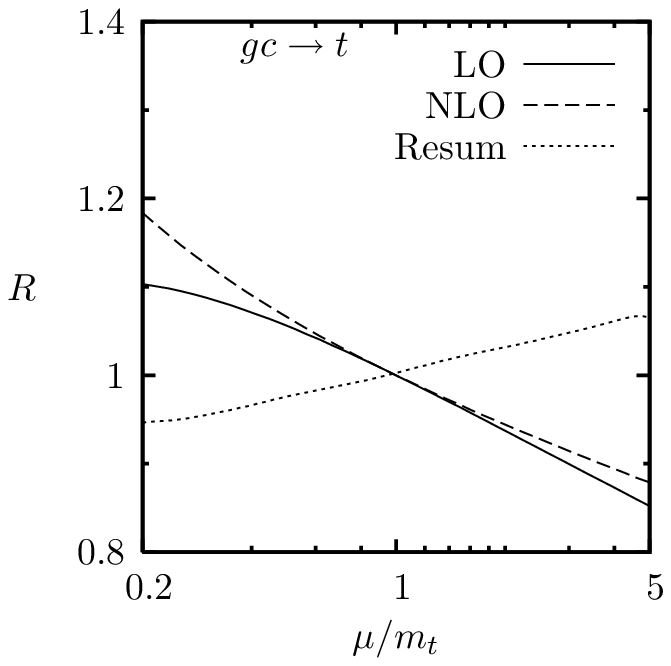}
  \caption{\label{fig:lhc_cteq}The ratio $R$ as functions of the renormalization and
    factorization scales at the LHC for subprocesses $gu \to t$ (left) and $gc \to t$
    (right) using CTEQ6 PDFs.}
\end{figure}
\begin{figure}[ht!]
  \includegraphics[width=0.23\textwidth]{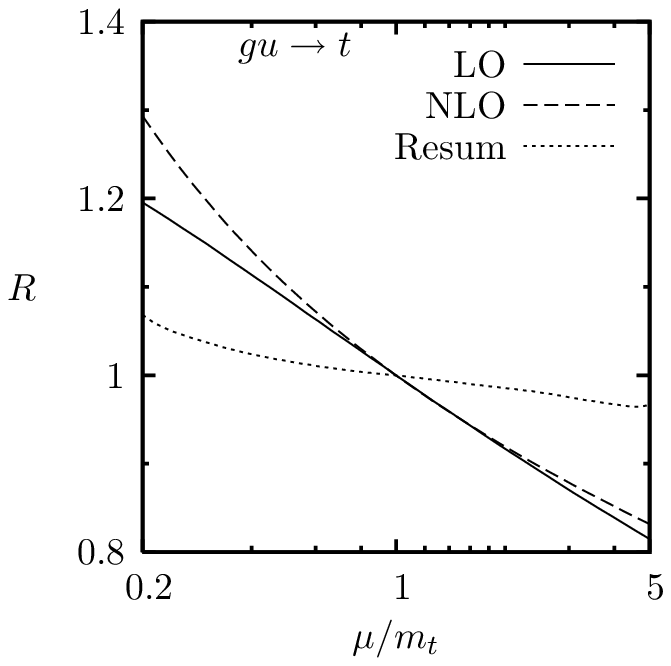}
  \includegraphics[width=0.23\textwidth]{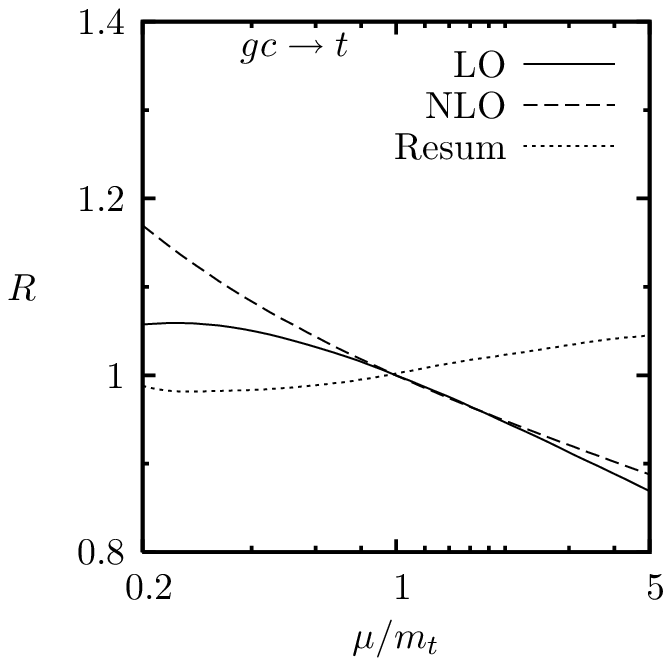}
  \caption{\label{fig:lhc_mrst}The ratio $R$ as functions of the renormalization and
    factorization scales at the LHC for subprocesses $gu \to t$ (left) and $gc \to t$
    (right) using MRST2004 PDFs.}
\end{figure}
In Figure~\ref{fig:lhc_cteq} and Figure~\ref{fig:lhc_mrst}, the ratio $R$ is shown for
the cases at the LHC. First, we note again that the NLO corrections do not reduce the
scale dependence of the LO cross sections. In fact, the variations of the LO and NLO
cross sections between $m_t/2$ and $2m_t$ are nearly the same: 18\% for $gu \to t$ and
12\% for $gc \to t$. After taking into account the threshold resummation effects, the
scale dependences are significantly reduced: the above variations become 5\% and 6\% for
$gu \to t$ and $gc \to t$, respectively. Especially, the resummed results have good
control over the scale dependences in the region $\mu < m_t$, where the NLO results
behave worse than the LO ones.

In summary, we have calculated the NLL threshold resummation effects in the direct top
quark productions induced by model-independent FCNC couplings at hadron colliders in the
framework of SCET and HQET. Our results show that the resummation effects increase the
total cross sections by about 30\% in general compared to the NLO results. Moreover, the
resummation effects significantly reduce the dependence of the cross sections on the
renormalization and factorization scales. Especially, the resummation effects have good
control over the scale dependences in the region $\mu < m_t$ at the LHC, where the NLO
results behave worse than the LO results. If future experiments at hadron colliders find
signals of the direct top quark production, the anomalous couplings can be extracted by
measuring the cross sections and comparing with the theoretical predictions. Thus, the
precision of the extracted anomalous couplings directly depends on the accuracy of the
theoretical predictions. Since the resummation effects increase the cross sections and
reduce the theoretical uncertainties, our results are more sensitive to the new physics
effects and it is more appropriate to use our results as the theoretical inputs.

\begin{acknowledgments}
  This work was supported in part by the National Natural Science Foundation of China,
  under grants No.~10421503 and No.~10575001, the Key Grant Project of Chinese Ministry
  of Education under grant No.~305001, and China Postdoctoral Science Foundation.
\end{acknowledgments}

\end{document}